**ABCMETAapp: R Shiny Application for Simulation-based Estimation of Mean and Standard Deviation for Meta-analysis via Approximate Bayesian Computation (ABC)**

Roopesh Reddy Sadashiva Reddy[1]
Email: roopesh@med.miami.edu

Isildinha M. Reis[1,2]
Email: ireis@med.miami.edu

Deukwoo Kwon[1,2]*
*Corresponding author
Email: DKwon@med.miami.edu

[1]Sylvester Comprehensive Cancer Center, University of Miami, Miami, FL 33136
[2]Department of Public Health Sciences, University of Miami, Miami, FL 33136

**Abstract**

*Background and Objective:* In meta-analysis based on continuous outcome, estimated means and corresponding standard deviations from the selected studies are key inputs to obtain a pooled estimate of the mean and its confidence interval. We often encounter the situation that these quantities are not directly reported in the literatures. Instead, other summary statistics are reported such as median, minimum, maximum, quartiles, and study sample size. Based on available summary statistics, we need to estimate estimates of mean and standard deviation for meta-analysis.

*Methods:* We developed a R Shiny code based on approximate Bayesian computation (ABC), ABCMETA, to deal with this situation.

*Results:* In this article, we present an interactive and user-friendly R Shiny application for implementing the proposed method (named ABCMETAapp). In ABCMETAapp, users can choose an underlying outcome distribution other than the normal distribution when the distribution of the outcome variable is skewed or heavy tailed. We show how to run ABCMETAapp with examples.

*Conclusions:* ABCMETAapp provides a R Shiny implementation. This method is more flexible than the existing analytical methods since estimation can be based on five different distribution (Normal, Lognormal, Exponential, Weibull, and Beta) for the outcome variable.

# 1. Introduction

Meta-analysis seeks to systematically review, summarize, integrate all pertinent evidence across published studies, and provide a more reliable pooled estimate of the outcome of interest. When we want to estimate the pooled mean of a continuous outcome, the main inputs for this analysis are estimated means and the corresponding standard deviations (or equivalently, variances) from the selected studies. In meta-analysis, we often encounter that some studies do not report sample means and sample standard deviations. Instead, other summary statistics are reported such as confidence interval, median, minimum value, maximum value, range or interquartile range (IQR).

Three existing approaches for estimating study specific mean and standard deviation were widely used. Hozo et al. [1] proposed a simple method for estimating the sample mean and the sample standard deviation from minimum value, median, maximum value, and the sample size. Bland [2] proposed another method based on minimum value, first quartile, median, third quartile, maximum value, and the sample size. Wan et al. [3] proposed a method covering above two situations for Hozo et al. [1] and Bland [2] and provided improved estimation method based on median, first quartile, third quartile, and the sample size. Since these three existing methods are easy to implement using Excel or R, their popularity is quite high. According to citation numbers from Scopus website (www.scopus.com), Hozo et al. [1] has 3,016 citations and both Bland [2] and Wan et al. [3] have same 869 citations as of February 27, 2020. Kwon and Reis [4] proposed a simulation-based approach using approximate Bayesian computation (ABC), ABCMETA. They showed ABCMETA produced more precise estimates of sample mean and sample standard deviation as sample size increases from the simulation study. One of the advantages of ABCMETA over the other existing methods is that ABCMETA allows estimation based on five distributions (Normal, Lognormal, Exponential, Weibull, and Beta distributions) for the continuous outcome. Although ABCMETA has advantages to estimate sample mean and sample standard deviation compared to the existing methods, its citation number is only 8 citations as of the same date. One main reason for low utilization comes from lack of implementation tool.

In the article, we present an interactive and user-friendly R Shiny application for implementing the proposed ABCMETA method (named ABCMETAapp). In Methods, we describe briefly the ABCMETA method. We describe the main functionalities of the Interactive R Shiny Application ABCMETAapp. In Results, we show examples illustrating how to implement ABCMETAapp. Finally, in Discussion, advantages of ABCMETAapp are shown and some future extension is discussed.

## 2. Methods

In a meta-analysis for estimating pooled mean and corresponding 95% confidence interval, inputs are sample means and sample standard deviations for the selected studies. When we do not have these quantities for some published studies, we need to estimate those using reported summary statistics in those publications. Kwon and Reis [4] proposed ABCMETA to estimate sample mean and sample standard deviation. In this section, we summarize briefly ABCMETA.

ABCMETA uses same sets of summary statistics like in Hozo et al. [1], Bland [2], and Wan et al. [3]. These are minimum value, first quartile, median, third quartile, maximum value, and the sample size. The ABCMETAapp allows us to enter inputs for three situations according to the available summary statistics in a selected study. The first situation indicates that minimum value, median, maximum value, and sample size are available. The second situation indicates additionally having estimates of the first and third quartiles along with minimum value, median, maximum value, and sample size. The third situation indicates that median, first quartile, third quartile, and sample size are available.

*2.1 ABCMETA: Simulation-based method via Approximate Bayesian Computation*

Approximate Bayesian Computation (ABC) is an appropriate method to estimate sample means and sample standard deviations using other reported summary statistics. For statistical inference for sample mean and sample standard deviation, we cannot evaluate likelihood function since we do not have all data points. Using ABC approach, the likelihood function can be replaced by a comparison of summary statistics from the observed data and those from simulated data using a distance measure. The ABC method was introduced by Tavaré et al. [5] using a simple rejection approach to avoid the likelihood function computation using a simulated data from a specific distribution. Beaumont [6] and Marin et al. [7] provide more detailed description of ABC method.

Kwon and Reis [4] proposed ABCMETA to estimate sample mean and standard deviation using available summary statistics. Figure 1 describes steps of ABC method, which is used in ABCMETA. **The first step** is to generate a set of candidate values for parameters, $\theta^*$, from a specific prior distribution, $p(\theta)$. **The second step** is to generate pseudo data, $D^*$, from the probability distribution, $f(\theta^*)$. **The third step** is to calculate summary statistics based on pseudo data, $D^*$. In this step, we also obtain sample mean and sample standard deviation from pseudo data, D*. **The fourth step** is to decide whether θ* are accepted or not. This decision is based on the distance between observed summary statistics, $S_{obs}$, and those of simulated data, $\hat{S}$, i.e., $\rho(S_{obs}, \hat{S})$, where $\rho(\bullet, \bullet)$ denotes a distance measure. The Euclidean distance is used in ABCMETA. If a distance between $S_{obs}$ and $\hat{S}$ is smaller than a pre-specified tolerance value, ε, (i.e., $\rho(S_{obs}, \hat{S}) < \varepsilon$) then θ* is accepted, otherwise it is rejected. At the same time, if θ* is accepted, calculated sample mean and sample standard deviation from pseudo data, D*, are also accepted. **Steps 1-4** are repeated many times (e.g., 100,000 repeats) to obtain enough number of θ*, sample mean, and sample standard deviation. The fundamental idea of ABC is that a good approximation of the posterior distribution can be obtained using summary statistics and a small tolerance value, ε, when likelihood function evaluation is infeasible due to absence of raw data or very expensive in computational cost. This approximation is represented in the form of posterior, $p(\theta | \rho(S_{obs}, \hat{S}) < \varepsilon)$. However, when we implement ABC method, it is difficult to decide appropriate tolerance value, ε, since it depends on unit of data. Instead of setting small value for ε, in our ABCMETAapp, we set the acceptance percentage. For example, acceptance percentage of 0.1% and 100,000 repeats mean that we retain 100 θ*'s corresponding to the top 0.1% shortest distances. Our recommendation for the acceptance percentage is 0.1%. However, a user can choose his/her own value such as 0.01% or 0.05% with much larger number of repeats.

In our ABCMETAapp, the first step is to choose one distribution among five underlying distributions of the continuous outcome to be used for generating pseudo data. The choice of underlying distribution is driven in part by whether the continuous variable is a bounded or unbounded. Typical example of bounded variable is some score of health-related quality of life (HRQoL) such as the University of Washington Quality of Life Questionnaire (UW-QOL) score ranging from 0 to 100, and a pain intensity score from a Visual Analogue Scale (VAS) ranging from 0 to 10. For a bounded outcome, the user of ABCMETAapp should choose as underlying distribution a beta distribution. Since user provides lower limit and upper limit for the continuous outcome, ABCMETAapp internally rescales the actual variable range to unit interval ([0, 1]) and report estimates of sample mean and standard deviation in original scale. For unbounded variable, we suggest to choose one from following four different distributions: normal distribution, lognormal distribution, exponential distribution, and Weibull distribution. If the continuous outcome can have a value between $-\infty$ and $\infty$, the natural choice for underlying distribution is the normal distribution (e.g., absolute differences (change) from baseline). If the outcome can have only non-negative values then we can consider lognormal, exponential, or Weibull distribution (e.g., pollutant concentrations in air). After deciding an underlying distribution, prior distributions for parameters are determined by corresponding chosen distribution in the previous step. If normal or log-normal distribution is chosen, location and scale parameters, $\mu$ and $\sigma$, are needed to specify. For Weibull distribution, shape and scale parameters are needed. For beta distribution, two shape parameters are needed. Usual choice of the priors is uniform distribution with relative wide range in ABC method. When a chosen distribution belong to location-scale family such as normal and log-normal distributions, we can use an educated guess for location parameter, $\mu$. Instead of uniform distribution with wide range, we can use given descriptive statistics such as minimum value (first quartile if it is available) for lower bound and maximum value (third quartile if it is available) for upper bound of uniform distribution. Other prior distributions for shape and scale parameters are uniform between zero and some large number. Table 1 shows recommended prior setting for each chosen underlying distribution. The estimates of mean and standard deviation from ABCMETAapp are obtained based on means of accepted values for mean and standard deviation.

*2.2 Interactive R Shiny Application (ABCMETAapp)*

ABCMETAapp is a R shiny program to implement ABCMETA approach. R code is available in Supplemental materials.

There are five steps to run ABCMETAapp in order to obtain the estimated mean and standard deviation. These are as follows:

1. **Data input**: Available summary statistics and sample size should be entered by the user. Input boxes are shown according to the chosen scenario of available summary statistics.

2. **Selection of the underlying distribution**: User chooses one distribution among five candidate distributions (Normal, Log-normal, Exponential, Weibull and Beta distributions).

3. **Determination of upper limit values for prior distributions**: Default upper limit values for prior distributions for parameters of the underlying distribution are shown in Table 1. The values can be change by the user.

4. **Total number of simulation and acceptance percentage**: default number of simulation and acceptance percentage are 50,000 and 0.1%. These values can be change by the user.

5. **Run application**: When user clicks 'Run ABCMETA' button, progress bar is displayed and estimates of sample mean and sample standard deviation are shown at top right position.

In Figure 2, we show the three input situations according to available summary statistics. Figure 2 (left) is for the situation that minimum value, median, maximum value, and sample size are available. Figure 2 (middle) is for the situation that median, first quartile, third quartile, and sample size are available. Figure 2 (right) is for the situation that the first and third quartiles along with minimum value, median, maximum value, and sample size. In Figure 3, we show the five underlying distributions and corresponding prior setting along with distribution selection. Users can change upper limit values for priors if necessary. The number of simulation for ABCMETA and acceptance percentage should be specified. Default values are automatically shown but users can change those values.

*2.3 Distribution Selection feature in ABCMETAapp*

When user wants to do selection for underlying distribution, ABCMETAapp provides estimates for mean and standard deviation from the distribution with highest selection probability. In this feature, we use default values for priors of parameters in Table 1. The distribution selection feature can be implemented by clicking check box for '**Distribution Selection**'. ABCMETAapp gives chosen underlying distribution and selection probability for that distribution as well as estimates of mean and standard deviation. Calculation for selection probability can be found in Kwon and Reis [4].

## 3. Results

We demonstrated ABCMETAapp with four hypothetical examples. We used sample size is 500 for all examples. The first example was that we have median, the first quartile, and third quartile for difference (first quartile=-1.4 median=-0.2, third quartile=0.95). Difference of 1.2 between first quartile and median is similar to difference of 0.75 between median and third quartile. Hence we chose normal distribution as underlying distribution. We obtained estimates of mean and standard deviation with 50,000 iterations and 0.1% acceptance percentage: mean=-0.225 and SD=1.744. We also conducted sensitivity analysis in terms of the number of iterations in ABCMETAapp. With 100,000 and 500,000 iterations we had mean=-0.22 and SD=1.745; and mean=-0.223 and SD=1.750, respectively.

The second example is hypothetical HRQoL score which has range between 0 and 100. We have the following summary statistics (minimum value=2.7, median=72.5, maximum value=99.9). Since nature of data is bounded, beta distribution was chosen as the underlying distribution. For this example, we ran 100,000 iterations and 0.1% acceptance percentage. Estimated mean and SD were 67.42 and 22.55.

The third example is the situation that summary statistics show some asymmetry. First we generate a random sample of 500 observations from a lognormal distribution with µ=1.5 and σ=0.5. These parameter values of lognormal distribution give us true mean=5.15 and SD=2.9. In the sample, we obtained three summary statistics (minimum value=0.82, median=4.44, maximum value=22.15). In this example, the difference of 3.62 between minimum value and median is much smaller than difference of 17.71 between median and maximum value. Normal distribution as an underlying distribution is not a good choice due to asymmetric nature of summary statistics. Therefore, we can choose one of three distributions (Lognormal, exponential, and Weibull distributions). We ran ABCMETAapp using 'Distribution Selection' feature with 100,000 iterations and 0.1% acceptance percentage. The probability of lognormal distribution was 0.66 and estimates of mean and SD from lognormal distribution were 4.932 and 2.95, respectively.

In the last example, we consider the situation that the summary statistics show some asymmetry with one or more summary statistics being a negative value. Obviously three distributions used in the previous example have non-negative support. Although normal distribution does not have issue about support, it cannot deal with asymmetry. For this situation, we have as an ad-hoc solution the addition of a constant *c* such that the available summary statistics become all positive. The addition of a constant does not change the estimate of standard deviation. For mean estimate, we obtain mean estimate by subtracting constant value *c* from output of ABCMETAapp. For example, suppose we have the following summary statistics minimum value=-9.65, median=-5.59, and maximum value=39.25. To make all values positive, we can add 10 to all summary statistics. Then new summary statistics, after adding the constant 10, are minimum value=0.35, median=4.41, and maximum value=49.25. Based on the new summary statistics, the estimates of mean and SD from ABCMETAapp were 6.67 and 6.84 from exponential distribution, respectively. The correct mean estimate, -3.33, is obtained by subtracting 10 from the estimated mean 6.67 of ABCMETAapp.

## 4. Discussion

Kwon and Reis [4] showed ABCMETA performs well in modest or large sample size. This method can handle skewed or heavy tailed distributions through distribution selection unlike Hozo et al. [1], Bland [2] and Wan et al. [3]. However, these three methods have been used widely due to easy implementation. This article provides an implementation tool for ABCMETA to obtain estimates of mean and standard deviation. ABCMETAapp makes Kwon and Reis [4] method publicly available. Many researchers can perform estimation of sample mean and sample standard deviation using methods described in this article to assess sensitivity of their results. The current implementation of ABCMETAapp is based on simple rejection method. We plan to implement other advanced approaches in ABC such as sequential Monte Carlo (ABC-SMC; Toni et al. [6]) to obtain more precise estimates of sample mean and sample standard deviation.

# Declaration of Competing Interest

None.

# Appendix

The appendix contains ABCMETAapp R code.

```
library(shiny)
library(shinyjs)

# Define the UI
ui <- fluidPage(

  # App title ----
  titlePanel("ABCMETAapp: Estimating Mean and Standard Deviation via ABC (Approximate Bayesian Computation)"),

  # Sidebar layout with input and output definitions ----
  sidebarLayout(
    sidebarPanel( style = "position:fixed;width:inherit;",width = 5.5,
            tags$head(tags$script('
                    var dimension = [0, 0];
                        $(document).on("shiny:connected", function(e) {
                        dimension[0] = window.innerWidth;
                        dimension[1] = window.innerHeight;
                        Shiny.onInputChange("dimension", dimension);
                        });
                        $(window).resize(function(e) {
                        dimension[0] = window.innerWidth;
                        dimension[1] = window.innerHeight;
                        Shiny.onInputChange("dimension", dimension);
                        });
                        ')),

      textInput("integer", label = "Sample Size(n):", value = "500" , width = '200px'),
      radioButtons("Scenarios", "Available summary statistics:",
              c("Min, Med, Max" = "sc1",
                "Q1, Med, Q3" = "sc2",
                "Min, Q1, Med, Q3, Max" = "sc3")),
      #condition for radiobutton sc1
      conditionalPanel(
        condition = "input.Scenarios == 'sc1'",
        textInput("ss1min", label = "Minimum:", value = "" , width = '200px'),
        textInput("ss1med", label = "Median:", value = "" , width = '200px'),
        textInput("ss1max", label = "Maximum:", value = "" , width = '200px')),
      #condition for radiobutton sc2
      conditionalPanel(
        condition = "input.Scenarios == 'sc2'",
        textInput("ss2q1", label = "1st Quartile:", value = "" , width = '200px'),
        textInput("ss2med", label = "Median:", value = "" , width = '200px'),
        textInput("ss2q3", label = "3rd Quartile:", value = "" , width = '200px')),
      #condition for radiobutton ss3
      conditionalPanel(
        condition = "input.Scenarios == 'sc3'",
        textInput("ss3min", label = "Minimum:", value = "" , width = '200px'),
        textInput("ss3q1", label = "1st Quartile:", value = "" , width = '200px'),
        textInput("ss3med", label = "Median:", value = "" , width = '200px'),
        textInput("ss3q3", label = "3rd Quartile:", value = "" , width = '200px'),
        textInput("ss3max", label = "Maximum:", value = "" , width = '200px')),
      selectInput("Typeofdistribution", label = "Underlying distribution:", c("Normal" = "normal","LogNormal" = "lognormal","Exponential" = "exponential","Weibull" = "weibull","Beta" = "beta", "Distribution Selection" = "DistributionSelection" ), width = "300px"),

      conditionalPanel(
        condition = "input.Typeofdistribution == 'normal'",
        textInput("nsigma", label = "Upper limit value for Sigma:", value = "50" , width = '200px')),
      conditionalPanel(
        condition = "input.Typeofdistribution == 'lognormal'",
        textInput("lsigma", label = "Upper limit value for Sigma:", value = "10" , width = '200px')),
      conditionalPanel(
```

```
                    condition = "input.Typeofdistribution == 'exponential'",
                    textInput("elambda", label = "Upper limit value for Lambda:", value = "40" , width = '200px')),
                 conditionalPanel(
                    condition = "input.Typeofdistribution == 'beta'",
                    textInput("lbbeta", label = "Lower bound value:", value = "0" , width = '200px'),
                    textInput("ubbeta", label = "Upper bound value:", value = "100" , width = '200px'),
                    textInput("balpha", label = "Upper limit value for Alpha:", value = "40" , width = '200px'),
                    textInput("bbeta", label = "Upper limit value for Beta:", value = "40" , width = '200px')),

                 conditionalPanel(
                    condition = "input.Typeofdistribution == 'weibull'",
                    textInput("wlambda", label = "Upper limit value for Lambda:", value = "50" , width = '200px'),
                    textInput("wkappa", label = "Upper limit value for Kappa:", value = "50" , width = '200px')),

                 ## setup for ABC
                 textInput("nb_simul", label = "Total No. of simulations:", value = "50000" , width = '200px'),   textInput("accperc", label =
"Acceptance percentage (%):", value = "0.1" , width = '200px'),  actionButton("goButton", "Run ABCMETA!")),

         mainPanel(column(9, offset = 8,verbatimTextOutput("nText")))

         ,position = c("left", "right")
    ) )

# Define the server code
server <- function(input, output, session) {

  observeEvent(input$Senarios, {
    updateTextInput(session, "ss1min",value = "");
    updateTextInput(session, "ss1med",value = "");
    updateTextInput(session, "ss1max",value = "");

    updateTextInput(session, "ss2q1",value = "");
    updateTextInput(session, "ss2med",value = "");
    updateTextInput(session, "ss2q3",value = "");

    updateTextInput(session, "ss3min",value = "");
    updateTextInput(session, "ss3q1",value = "");
    updateTextInput(session, "ss3med",value = "");
    updateTextInput(session, "ss3q3",value = "");
    updateTextInput(session, "ss3max",value = "");

  })

  ntext <- eventReactive(input$goButton,
                         {

                             nb_simul.val= as.numeric(input$nb_simul);
                             acc.perc= as.numeric(input$accperc)/100;
                             random.seed=1234;
                             distrib= input$Typeofdistribution
                             set.seed(random.seed);
                             n=input$integer;

                             progress <- shiny::Progress$new()
                             on.exit(progress$close())
                             progress$set(message = "Simulate", value = 0)

                             if(input$Typeofdistribution == 'normal')
                             {sigma = as.numeric(input$nsigma);}

                             if(input$Typeofdistribution == 'lognormal')
                             {sigma = as.numeric(input$lsigma);}

                             if(input$Typeofdistribution == 'exponential')
                             {lambda = as.numeric(input$elambda);}

                             if(input$Typeofdistribution == 'beta')
```

```r
              {alpha = as.numeric(input$balpha);
               beta = as.numeric(input$bbeta);
               limit1 = as.numeric(input$lbbeta);
               limit2 = as.numeric(input$ubbeta);
               }

              if(input$Typeofdistribution == 'weibull')
              {lambda = as.numeric(input$wlambda);
               kappa = as.numeric(input$wkappa);
               }

              if(input$Scenarios == 'sc1'){summ.val<-
c(as.numeric(input$ss1min),NA,as.numeric(input$ss1med),NA,as.numeric(input$ss1max));}

              if(input$Scenarios == 'sc2'){summ.val<-
c(NA,as.numeric(input$ss2q1),as.numeric(input$ss2med),as.numeric(input$ss2q3),NA);}

              if(input$Scenarios == 'sc3'){summ.val<-
c(as.numeric(input$ss3min),as.numeric(input$ss3q1),as.numeric(input$ss3med),as.numeric(input$ss3q3),as.numeric(input$ss3max));}

              up.ind= as.numeric(nb_simul.val)*as.numeric(acc.perc); ## 50 (top 0.1% among 50000);

              # Matrix of simulated mean and sd values
              est.mat=matrix(NA,ncol=10,nrow=nb_simul.val)
              # distance matrix (1:normal, 2:lognormal, 3:exponential, 4:weibull, 5:beta)
              dist.mat= matrix(NA,ncol=5,nrow=nb_simul.val)

##########################################################################

              if (distrib=='normal'){

                for (i in 1:nb_simul.val){

                if (is.na(summ.val[2])==FALSE & is.na(summ.val[4])==FALSE){mustar1=runif(1,summ.val[2],summ.val[4]); }
                if (is.na(summ.val[2])==TRUE & is.na(summ.val[4])==TRUE){mustar1=runif(1,summ.val[1],summ.val[5]); }

                sigstar1=runif(1,0,sigma);
                temp.sam=rnorm(n,mustar1,sigstar1);
                est.mat[i,1]=mean(temp.sam);
                est.mat[i,2]=sd(temp.sam);

                ss1=min(temp.sam);ss2=quantile(temp.sam,.25)[[1]];
                ss3=median(temp.sam);ss4= quantile(temp.sam,.75)[[1]];
                ss5=max(temp.sam);

                if (is.na(summ.val[2])==TRUE & is.na(summ.val[4])==TRUE){dist.mat
[i,1]=sqrt(sum((c(summ.val[1],summ.val[3],summ.val[5])-c(ss1,ss3,ss5))^2));}
                if (is.na(summ.val[1])==TRUE & is.na(summ.val[5])==TRUE){dist.mat
[i,1]=sqrt(sum((c(summ.val[2],summ.val[3],summ.val[4])-c(ss2,ss3,ss4))^2));}
                if (is.na(summ.val[1])==FALSE & is.na(summ.val[2])==FALSE & is.na(summ.val[4])==FALSE &
is.na(summ.val[5])==FALSE){dist.mat [i,1]=sqrt(sum((c(summ.val[1],summ.val[2],summ.val[3],summ.val[4],summ.val[5])-
c(ss1,ss2,ss3,ss4,ss5))^2));}

                if(i%%500==0){print(paste('iter======',i));
                  progress$inc(500/(nb_simul.val), detail = paste("Doing part", i, " of ",nb_simul.val))
                }
      } # end of sim

                ind=sort(dist.mat[,1],index.return=T)$ix;mycol=1:2;
                output=est.mat[ind[1:up.ind],mycol]
                est= apply(output,2,'mean')

              } # end of normal
```

############################################################

```r
if (distrib=='lognormal'){

  for (i in 1:nb_simul.val){

    if (is.na(summ.val[2])==FALSE & 
is.na(summ.val[4])==FALSE){mustar2=runif(1,log(summ.val[2]),log(summ.val[4])); }
    if (is.na(summ.val[2])==TRUE & is.na(summ.val[4])==TRUE){mustar2=runif(1,log(summ.val[1]),log(summ.val[5])); }

    sigstar2=runif(1,0,sigma);
    temp.sam=rlnorm(n,mustar2,sigstar2);
    est.mat[i,3]=mean(temp.sam);
    est.mat[i,4]=sd(temp.sam);
    ss1=min(temp.sam);ss2=quantile(temp.sam,.25)[[1]];
    ss3=median(temp.sam);ss4= quantile(temp.sam,.75)[[1]];
    ss5=max(temp.sam);

    if (is.na(summ.val[2])==TRUE & is.na(summ.val[4])==TRUE){dist.mat
[i,2]=sqrt(sum((c(summ.val[1],summ.val[3],summ.val[5])-c(ss1,ss3,ss5))^2));}
    if (is.na(summ.val[1])==TRUE & is.na(summ.val[5])==TRUE){dist.mat
[i,2]=sqrt(sum((c(summ.val[2],summ.val[3],summ.val[4])-c(ss2,ss3,ss4))^2));}
    if (is.na(summ.val[1])==FALSE & is.na(summ.val[2])==FALSE & is.na(summ.val[4])==FALSE & 
is.na(summ.val[5])==FALSE){dist.mat [i,2]=sqrt(sum((c(summ.val[1],summ.val[2],summ.val[3],summ.val[4],summ.val[5])-
c(ss1,ss2,ss3,ss4,ss5))^2));}

    if(i%%500==0){print(paste('iter=======',i));
    progress$inc(500/(nb_simul.val), detail = paste("Doing part", i, " of ",nb_simul.val));}

  } # end of sim

  ind=sort(dist.mat[,2],index.return=T)$ix;mycol=3:4;

  output=est.mat[ind[1:up.ind],mycol]
  est= apply(output,2,'mean')

} # end of lognormal

if (distrib=='exponential'){

  for (i in 1:nb_simul.val){

    lamstar=runif(1,0,lambda);
    temp.sam=rexp(n,1/lamstar);
    est.mat[i,5]=mean(temp.sam);
    est.mat[i,6]=sd(temp.sam);
    ss1=min(temp.sam);ss2=quantile(temp.sam,.25)[[1]];
    ss3=median(temp.sam);ss4= quantile(temp.sam,.75)[[1]];
    ss5=max(temp.sam);

    if (is.na(summ.val[2])==TRUE & is.na(summ.val[4])==TRUE){dist.mat
[i,3]=sqrt(sum((c(summ.val[1],summ.val[3],summ.val[5])-c(ss1,ss3,ss5))^2));}
    if (is.na(summ.val[1])==TRUE & is.na(summ.val[5])==TRUE){dist.mat
[i,3]=sqrt(sum((c(summ.val[2],summ.val[3],summ.val[4])-c(ss2,ss3,ss4))^2));}
    if (is.na(summ.val[1])==FALSE & is.na(summ.val[2])==FALSE & is.na(summ.val[4])==FALSE & 
is.na(summ.val[5])==FALSE){dist.mat [i,3]=sqrt(sum((c(summ.val[1],summ.val[2],summ.val[3],summ.val[4],summ.val[5])-
c(ss1,ss2,ss3,ss4,ss5))^2));}
```

```
            if(i%%500==0){print(paste('iter=======',i));
            progress$inc(500/(nb_simul.val), detail = paste("Doing part", i, " of ",nb_simul.val))
            }

        } # end of sim
        ind=sort(dist.mat[,3],index.return=T)$ix;mycol=5:6;
        output=est.mat[ind[1:up.ind],mycol]
        est= apply(output,2,'mean')

      } # end of expo

##############################################################

        if (distrib=='weibull'){

          for (i in 1:nb_simul.val){

          lamstar=runif(1,0,lambda);
          kappastar=runif(1,0,kappa);
          temp.sam=rweibull(n,lamstar,kappastar);
          est.mat[i,7]=mean(temp.sam);
          est.mat[i,8]=sd(temp.sam);
          ss1=min(temp.sam);ss2=quantile(temp.sam,.25)[[1]];
          ss3=median(temp.sam);ss4= quantile(temp.sam,.75)[[1]];
          ss5=max(temp.sam);

            if (is.na(summ.val[2])==TRUE & is.na(summ.val[4])==TRUE){dist.mat
[i,4]=sqrt(sum((c(summ.val[1],summ.val[3],summ.val[5])-c(ss1,ss3,ss5))^2));}
            if (is.na(summ.val[1])==TRUE & is.na(summ.val[5])==TRUE){dist.mat
[i,4]=sqrt(sum((c(summ.val[2],summ.val[3],summ.val[4])-c(ss2,ss3,ss4))^2));}
            if (is.na(summ.val[1])==FALSE & is.na(summ.val[2])==FALSE & is.na(summ.val[4])==FALSE &
is.na(summ.val[5])==FALSE){dist.mat [i,4]=sqrt(sum((c(summ.val[1],summ.val[2],summ.val[3],summ.val[4],summ.val[5])-
c(ss1,ss2,ss3,ss4,ss5))^2));}

            if(i%%500==0){print(paste('iter=======',i));
              progress$inc(500/(nb_simul.val), detail = paste("Doing part", i, " of ",nb_simul.val))
              }

          } # end of sim
          ind=sort(dist.mat[,4],index.return=T)$ix;mycol=7:8;
          output=est.mat[ind[1:up.ind],mycol]
          est= apply(output,2,'mean')

        } # end of weibull

###########################################################
        if (distrib=='beta'){

          summ.val=(summ.val-limit1)/(limit2-limit1);
          difflim=(limit2-limit1);

          for (i in 1:nb_simul.val){

            alphastar=runif(1,0,alpha);
```

```r
              betastar=runif(1,0,beta);
              temp.sam=rbeta(n,alphastar,betastar);
              est.mat[i,9]=difflim*mean(temp.sam)+limit1;
              est.mat[i,10]=difflim*sd(temp.sam);

            ss1=min(temp.sam);ss2=quantile(temp.sam,.25)[[1]];
            ss3=median(temp.sam);ss4= quantile(temp.sam,.75)[[1]];
            ss5=max(temp.sam);

            if (is.na(summ.val[2])==TRUE & is.na(summ.val[4])==TRUE){dist.mat
[i,5]=sqrt(sum((c(summ.val[1],summ.val[3],summ.val[5])-c(ss1,ss3,ss5))^2));}
            if (is.na(summ.val[1])==TRUE & is.na(summ.val[5])==TRUE){dist.mat
[i,5]=sqrt(sum((c(summ.val[2],summ.val[3],summ.val[4])-c(ss2,ss3,ss4))^2));}
            if (is.na(summ.val[1])==FALSE & is.na(summ.val[2])==FALSE & is.na(summ.val[4])==FALSE &
is.na(summ.val[5])==FALSE){dist.mat [i,5]=sqrt(sum((c(summ.val[1],summ.val[2],summ.val[3],summ.val[4],summ.val[5])-
c(ss1,ss2,ss3,ss4,ss5))^2));}

            if(i%%500==0){print(paste('iter=======',i));
             progress$inc(500/(nb_simul.val), detail = paste("Doing part", i, " of ",nb_simul.val))
             }

          } # end of sim

          ind=sort(dist.mat[,5],index.return=T)$ix;mycol=9:10;

          output=est.mat[ind[1:up.ind],mycol]
          est= apply(output,2,'mean')

        } # end of beta

      if (distrib=='DistributionSelection'){

        set.seed(random.seed);
        up.ind= nb_simul.val*acc.perc; ## 50 (top 0.1% among 50000);

        # Matrix of simulated mean and sd values
        est.mat=matrix(NA,ncol=10,nrow=nb_simul.val)

        # distance matrix (1:normal, 2:lognormal, 3:exponential, 4:weibull, 5:beta)
        dist.mat= matrix(NA,ncol=5,nrow=nb_simul.val)

        for (i in 1:nb_simul.val){

          # normal distrib part

          if (is.na(summ.val[2])==FALSE & is.na(summ.val[4])==FALSE){mustar1=runif(1,summ.val[2],summ.val[4]); }
          if (is.na(summ.val[2])==TRUE & is.na(summ.val[4])==TRUE){mustar1=runif(1,summ.val[1],summ.val[5]); }

          sigstar1=runif(1,0,50);
          temp.sam=rnorm(n,mustar1,sigstar1);
          est.mat[i,1]=mean(temp.sam);
          est.mat[i,2]=sd(temp.sam);

          ss1=min(temp.sam);ss2=quantile(temp.sam,.25)[[1]];
          ss3=median(temp.sam);ss4= quantile(temp.sam,.75)[[1]];
          ss5=max(temp.sam);

          if (is.na(summ.val[2])==TRUE &
is.na(summ.val[4])==TRUE){dist.mat[i,1]=sqrt(sum((c(summ.val[1],summ.val[3],summ.val[5])-c(ss1,ss3,ss5))^2));}
          if (is.na(summ.val[1])==TRUE &
is.na(summ.val[5])==TRUE){dist.mat[i,1]=sqrt(sum((c(summ.val[2],summ.val[3],summ.val[4])-c(ss2,ss3,ss4))^2));}
```

```r
			if (is.na(summ.val[1])==FALSE & is.na(summ.val[2])==FALSE & is.na(summ.val[4])==FALSE & is.na(summ.val[5])==FALSE){dist.mat[i,1]=sqrt(sum((c(summ.val[1],summ.val[2],summ.val[3],summ.val[4],summ.val[5])-c(ss1,ss2,ss3,ss4,ss5))^2));}

			# lognormal distrib part

			if (is.na(summ.val[2])==FALSE & is.na(summ.val[4])==FALSE){mustar2=runif(1,log(summ.val[2]),log(summ.val[4])); }
			if (is.na(summ.val[2])==TRUE & is.na(summ.val[4])==TRUE){mustar2=runif(1,log(summ.val[1]),log(summ.val[5])); }

			sigstar2=runif(1,0,10);
			temp.sam=rlnorm(n,mustar2,sigstar2);
			est.mat[i,3]=mean(temp.sam);
			est.mat[i,4]=sd(temp.sam);
			ss1=min(temp.sam);ss2=quantile(temp.sam,.25)[[1]];
			ss3=median(temp.sam);ss4= quantile(temp.sam,.75)[[1]];
			ss5=max(temp.sam);

			if (is.na(summ.val[2])==TRUE & is.na(summ.val[4])==TRUE){dist.mat[i,2]=sqrt(sum((c(summ.val[1],summ.val[3],summ.val[5])-c(ss1,ss3,ss5))^2));}
			if (is.na(summ.val[1])==TRUE & is.na(summ.val[5])==TRUE){dist.mat[i,2]=sqrt(sum((c(summ.val[2],summ.val[3],summ.val[4])-c(ss2,ss3,ss4))^2));}
			if (is.na(summ.val[1])==FALSE & is.na(summ.val[2])==FALSE & is.na(summ.val[4])==FALSE & is.na(summ.val[5])==FALSE){dist.mat[i,2]=sqrt(sum((c(summ.val[1],summ.val[2],summ.val[3],summ.val[4],summ.val[5])-c(ss1,ss2,ss3,ss4,ss5))^2));}

			# exponential

			lamstar=runif(1,0,40);
			temp.sam=rexp(n,1/lamstar);
			est.mat[i,5]=mean(temp.sam);
			est.mat[i,6]=sd(temp.sam);
			ss1=min(temp.sam);ss2=quantile(temp.sam,.25)[[1]];
			ss3=median(temp.sam);ss4= quantile(temp.sam,.75)[[1]];
			ss5=max(temp.sam);

			if (is.na(summ.val[2])==TRUE & is.na(summ.val[4])==TRUE){dist.mat[i,3]=sqrt(sum((c(summ.val[1],summ.val[3],summ.val[5])-c(ss1,ss3,ss5))^2));}
			if (is.na(summ.val[1])==TRUE & is.na(summ.val[5])==TRUE){dist.mat[i,3]=sqrt(sum((c(summ.val[2],summ.val[3],summ.val[4])-c(ss2,ss3,ss4))^2));}
			if (is.na(summ.val[1])==FALSE & is.na(summ.val[2])==FALSE & is.na(summ.val[4])==FALSE & is.na(summ.val[5])==FALSE){dist.mat[i,3]=sqrt(sum((c(summ.val[1],summ.val[2],summ.val[3],summ.val[4],summ.val[5])-c(ss1,ss2,ss3,ss4,ss5))^2));}

			#   weibull'

			lamstar=runif(1,0,50);
			kappastar=runif(1,0,50);
			temp.sam=rweibull(n,lamstar,kappastar);
			est.mat[i,7]=mean(temp.sam);
			est.mat[i,8]=sd(temp.sam);
			ss1=min(temp.sam);ss2=quantile(temp.sam,.25)[[1]];
			ss3=median(temp.sam);ss4= quantile(temp.sam,.75)[[1]];
			ss5=max(temp.sam);

			if (is.na(summ.val[2])==TRUE & is.na(summ.val[4])==TRUE){dist.mat[i,4]=sqrt(sum((c(summ.val[1],summ.val[3],summ.val[5])-c(ss1,ss3,ss5))^2));}
			if (is.na(summ.val[1])==TRUE & is.na(summ.val[5])==TRUE){dist.mat[i,4]=sqrt(sum((c(summ.val[2],summ.val[3],summ.val[4])-c(ss2,ss3,ss4))^2));}
			if (is.na(summ.val[1])==FALSE & is.na(summ.val[2])==FALSE & is.na(summ.val[4])==FALSE & is.na(summ.val[5])==FALSE){dist.mat[i,4]=sqrt(sum((c(summ.val[1],summ.val[2],summ.val[3],summ.val[4],summ.val[5])-c(ss1,ss2,ss3,ss4,ss5))^2));}

			if(i%%500==0){print(paste('iter======',i));
			  progress$inc(500/(nb_simul.val), detail = paste("Doing part", i, " of ",nb_simul.val))
			} }

			dist=c(dist.mat[,1], dist.mat[,2], dist.mat[,3], dist.mat[,4]);
```

```
                model.ind=c(rep(1, nb_simul.val), rep(2, nb_simul.val), rep(3, nb_simul.val), rep(4, nb_simul.val));
                
                ind.all=sort(dist,index.return=T)$ix
                tb=rep(0,4);
                for (k in 1:4){tb[k]=length(which(model.ind[ind.all[1:up.ind]]==k));}
                
                sel.model= which(tb==max(tb))[1]
                if (sel.model==1){ind=sort(dist.mat[,1],index.return=T)$ix;mycol=1:2;sel.model.label='Normal';}
                if (sel.model==2){ind=sort(dist.mat[,2],index.return=T)$ix;mycol=3:4;sel.model.label='Log-Normal';}
                if (sel.model==3){ind=sort(dist.mat[,3],index.return=T)$ix;mycol=5:6;sel.model.label='Exponential';}
                if (sel.model==4){ind=sort(dist.mat[,4],index.return=T)$ix;mycol=7:8;sel.model.label='Weibull';}
                
                output=est.mat[ind[1:up.ind],mycol];
                est= apply(output,2,'mean');
                
                
              }
              
              if (distrib=='DistributionSelection'){
                paste("[","ABC Mean=",round(est[1],3),"]","[","ABC
SD=",round(est[2],3),"]","[","Distribution=",sel.model.label,"]","[","model prob=",tb[sel.model]/up.ind,"]");
              }
              else {paste("[","ABC Mean=",round(est[1],3),"]","[","ABC SD=",round(est[2],3),"]");}
              
  })
  
  output$nText <- renderText({ntext()});
  
  
}

# Return a Shiny app object
shinyApp(ui = ui, server = server, options = list(launch.browser = T ))
```

Table 1: Default priors for ABCMETA

| Distribution | Parameter 1 | Prior distribution for parameter 1 | Parameter 2 | Prior for parameter 2 |
|---|---|---|---|---|
| Normal (S1) | $\mu$ | Uniform ($X_{min}$, $X_{max}$) | $\sigma$ | Uniform(0,50) |
| Normal (S2) | $\mu$ | Uniform ($X_{Q1}$, $X_{Q3}$) | $\sigma$ | Uniform(0,50) |
| Normal (S3) | $\mu$ | Uniform ($X_{Q1}$, $X_{Q3}$) | $\sigma$ | Uniform(0,50) |
| Log-normal (S1) | $\mu$ | Uniform (log($X_{min}$), log($X_{max}$)) | $\sigma$ | Uniform(0,10) |
| Log-normal (S2) | $\mu$ | Uniform (log($X_{Q1}$), log($X_{Q3}$)) | $\sigma$ | Uniform(0,10) |
| Log-normal (S3) | $\mu$ | Uniform (log($X_{Q1}$), log($X_{Q3}$)) | $\sigma$ | Uniform(0,10) |
| Exponential | $\lambda$ | Uniform(0,40) | - | - |
| Beta | $\alpha$ | Uniform(0,40) | $\beta$ | Uniform(0,40) |
| Weibull | $\lambda$ | Uniform(0,50) | $\kappa$ | Uniform(0,50) |

Figure 1: Generic ABC steps

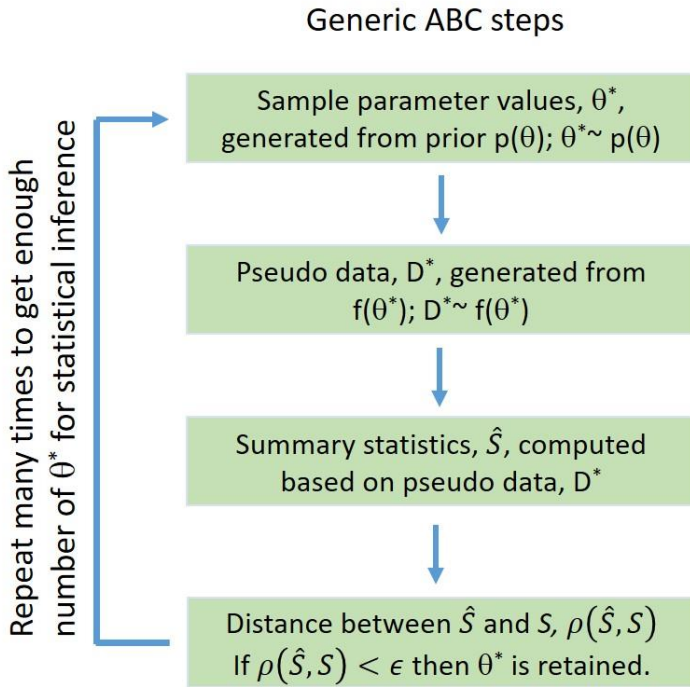

Figure 2: Three potential situations for ABCMETAapp

Figure 3: Parameter and running setup for ABCMETAapp for each underlying distribution